\documentclass[conference]{IEEEtran}
\IEEEoverridecommandlockouts
\pdfoutput=1
\ifCLASSINFOpdf
\else
\fi
\usepackage{makecell}

\usepackage{amsmath}
\usepackage{listings}
\usepackage{xcolor}
\usepackage{verbatim}
\usepackage{multirow}
\usepackage{graphicx}
\graphicspath{{./Figs/}}
\usepackage{url}
\usepackage{wasysym}
\usepackage{textcomp}
\usepackage{subcaption}
\usepackage{amssymb}
\usepackage{cite}
\usepackage{booktabs}
\usepackage{longtable}

\usepackage{array}
\usepackage{rotating}
\newcommand{\PreserveBackslash}[1]{\let\temp=\\#1\let\\=\temp}
\newcolumntype{C}[1]{>{\PreserveBackslash\centering}p{#1}}
\newcolumntype{R}[1]{>{\PreserveBackslash\raggedleft}p{#1}}
\newcolumntype{L}[1]{>{\PreserveBackslash\raggedright}p{#1}}

\usepackage[linesnumbered,lined, noend, algoruled,norelsize]{algorithm2e}
\usepackage{pifont}
\usepackage{mathtools}
\usepackage[colorlinks,linkcolor=blue,anchorcolor=blue,citecolor=blue]{hyperref}
\usepackage{flushend}
\usepackage{ulem}
\usepackage{algorithm2e}
\usepackage{seqsplit}
\SetAlFnt{\small}
\SetAlCapFnt{\small}
\SetAlCapNameFnt{\small}
\begin{document}
%

\title{Security Analysis of DeFi: Vulnerabilities, Attacks and Advances}
\author{
	\IEEEauthorblockN{
		Wenkai Li\thanks{\IEEEauthorrefmark{1} Wenkai Li and Jiuyang Bu are co-first authors.}\IEEEauthorrefmark{1},
		Jiuyang Bu\IEEEauthorrefmark{1},
		Xiaoqi Li\thanks{\IEEEauthorrefmark{2} The corresponding author.}\IEEEauthorrefmark{2},
		Xianyi Chen
	}
	\IEEEauthorblockA{
		School of Cyberspace Security, Hainan University, Haikou, China\\
		Email: liwenkai871@gmail.com, 736011577@qq.com, csxqli@gmail.com, 1550615836@qq.com}
}

\definecolor{verylightgray}{rgb}{.97,.97,.97}

\lstdefinelanguage{Solidity}{
	keywords=[1]{anonymous, assembly, assert, balance, break, call, callcode, case, catch, class, constant, continue, constructor, contract, debugger, default, delegatecall, delete, do, else, emit, event, experimental, export, external, false, finally, for, function, gas, if, implements, import, in, indexed, instanceof, interface, internal, is, length, library, log0, log1, log2, log3, log4, memory, modifier, new, payable, pragma, private, protected, public, pure, push, require, return, returns, revert, selfdestruct, send, solidity, storage, struct, suicide, super, switch, then, this, throw, transfer, true, try, typeof, using, value, view, while, with, addmod, ecrecover, keccak256, mulmod, ripemd160, sha256, sha3}, 
	keywordstyle=[1]\color{blue}\bfseries,
	keywords=[2]{address, bool, byte, bytes, bytes1, bytes2, bytes3, bytes4, bytes5, bytes6, bytes7, bytes8, bytes9, bytes10, bytes11, bytes12, bytes13, bytes14, bytes15, bytes16, bytes17, bytes18, bytes19, bytes20, bytes21, bytes22, bytes23, bytes24, bytes25, bytes26, bytes27, bytes28, bytes29, bytes30, bytes31, bytes32, enum, int, int8, int16, int24, int32, int40, int48, int56, int64, int72, int80, int88, int96, int104, int112, int120, int128, int136, int144, int152, int160, int168, int176, int184, int192, int200, int208, int216, int224, int232, int240, int248, int256, mapping, string, uint, uint8, uint16, uint24, uint32, uint40, uint48, uint56, uint64, uint72, uint80, uint88, uint96, uint104, uint112, uint120, uint128, uint136, uint144, uint152, uint160, uint168, uint176, uint184, uint192, uint200, uint208, uint216, uint224, uint232, uint240, uint248, uint256, var, void, ether, finney, szabo, wei, days, hours, minutes, seconds, weeks, years},	
	keywordstyle=[2]\color{teal}\bfseries,
	keywords=[3]{block, blockhash, coinbase, difficulty, gaslimit, number, timestamp, msg, data, gas, sender, sig, value, now, tx, gasprice, origin},	
	keywordstyle=[3]\color{violet}\bfseries,
	identifierstyle=\color{black},
	sensitive=false,
	comment=[l]{//},
	morecomment=[s]{/*}{*/},
	commentstyle=\color{gray}\ttfamily,
	stringstyle=\color{red}\ttfamily,
	morestring=[b]',
	morestring=[b]"
}

\lstset{
	language=Solidity,
	extendedchars=true,
	basicstyle=\footnotesize\ttfamily,
	showstringspaces=false,
	showspaces=false,
	numbers=left,
	numberstyle=\footnotesize,
	numbersep=9pt,
	tabsize=2,
	breaklines=true,
	showtabs=false,
	captionpos=b
}

\maketitle

\begin{abstract}
\label{abs}Decentralized finance (DeFi) in Ethereum is a financial ecosystem built on the blockchain that has locked over 200 billion USD until April 2022. All transaction information is transparent and open when transacting through the DeFi protocol, which has led to a series of attacks. Several studies have attempted to optimize it from both economic and technical perspectives. However, few works analyze the vulnerabilities and optimizations of the entire DeFi system. In this paper, we first systematically analyze vulnerabilities related to DeFi in Ethereum at several levels, then we investigate real-world attacks. Finally, we summarize the achievements of DeFi optimization and provide some future directions.
\end{abstract}


\section{Introduction}
\vspace{1ex}




The popularity of blockchain 2.0 technology has resulted in a wide range of related services. Decentralized finance (DeFi) is an example of a financial service built on blockchains to provide transaction transparency. From January 2020 to April 2022, the total value locked in DeFi climbs from 600 million USD to around 200 billion USD \cite{defillama}. However, there was a sharp drop in May 2022, which caused us to ponder the safety of DeFi system.

Attacks have emerged gradually with the rapid development of DeFi. Security incidents against DeFi continue to proliferate, and there has been a lot of research to improve the security of blockchain\cite{chen2018towards,chen2017adaptive,RN90,li2021hybrid,RN37,RN47,RN61,RN51,qin2021cefi,RN87}. \cite{RN90} presented \textsc{BLOCKEYE}, a real-time threat detection solution for Ethereum-based DeFi deployments. \cite{RN37} presented an online framework \textsc{SODA} to identify assaults on smart contracts. However, none of them consider economic security beyond detecting vulnerabilities against technical aspects. \cite{RN47} distinguished between technological and economic security and demonstrated that economic security is not flawless. \cite{RN61} investigated the scope of loan marketplaces and assess the risk of lending agreements. \cite{RN51} investigated how errors in design and pricing volatility of DeFi protocols might lead to DeFi crises. \cite{qin2021cefi} analyzed the differences between centralized finance (CeFi) and DeFi, covering legislation, economy, security, privacy, and market manipulation.

\begin{figure}[ht]
    \centering
    \includegraphics[width=8cm]{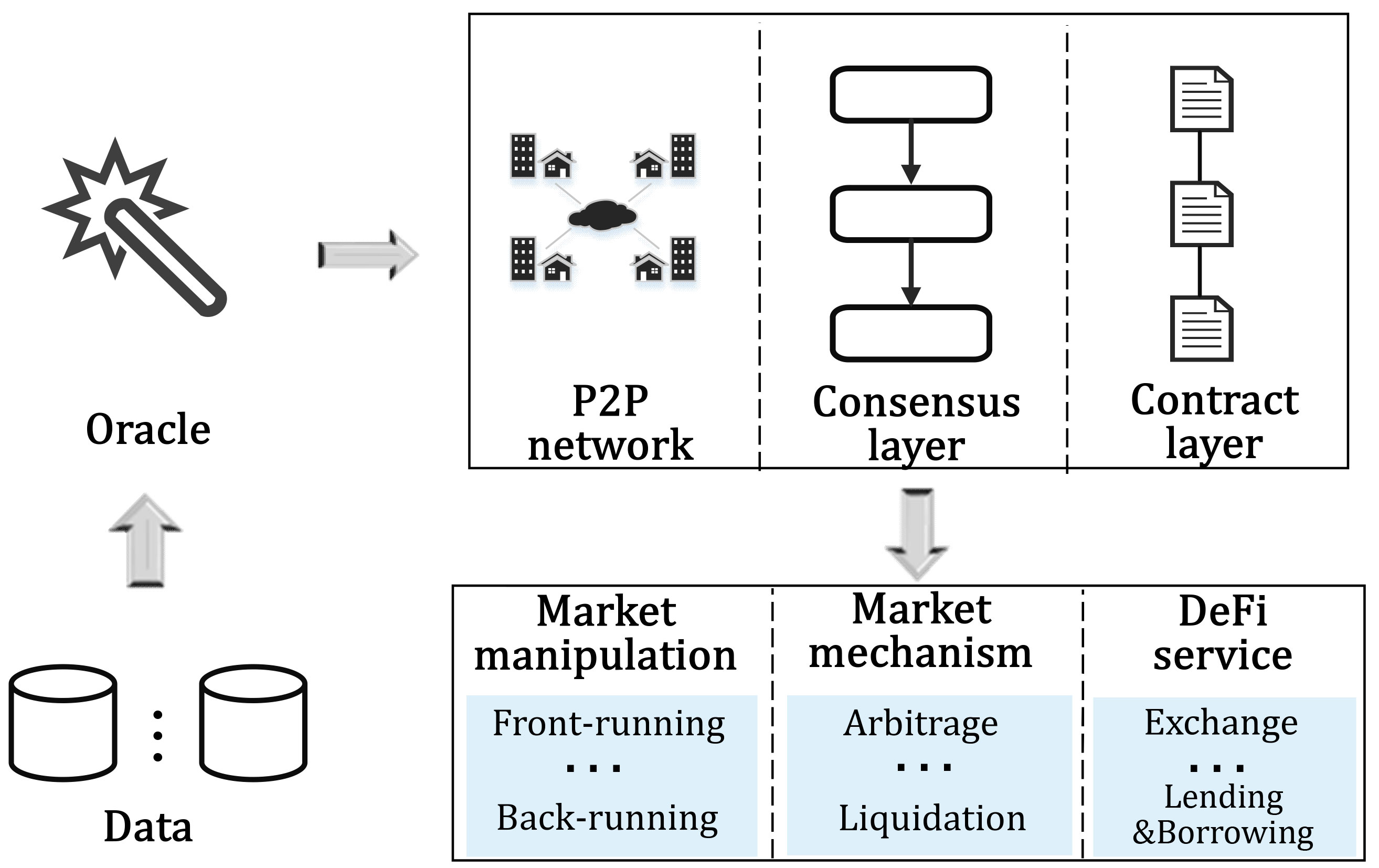}
    \caption{Overview of the Analysis Path}
    \label{fig:overview}
\end{figure}

However, DeFi is not secure enough that attacks on it keep appearing, such as the Ronin Bridge incident \cite{RN23}. It is shown in \cite{RN87} that the existing defenses do not reduce the number of attacks. Therefore, the security against DeFi needs to strengthen.

Unfortunately, there is still a lack of systematic analysis of DeFi system security. To assist in subsequent studies, our research analyzes the technical and economic risks to which Defi is vulnerable at the system level and comprehensively follows the path in Fig. \ref{fig:overview}. We also summarize existing real-world attacks in a way that provides a good foundation for future research, then we summarize classical related protection techniques. Finally, we provide an outlook on the possible improvements that need to strengthen in this area.

The main contributions of this paper are as follows:
\begin{itemize}
    \item [(1)] To the best of our knowledge, we conduct the \textit{first} systematic examination on the security issues of the DeFi ecosystem built on blockchain.

    \item [(2)] We systematically summarize the vulnerabilities of the Ethereum-based DeFi system, investigate real-world attack events related to DeFi and classify them according to their vulnerability principles.

    \item [(3)] We survey the security optimizations in DeFi from the system level and give some suggestions for future research directions in this area.
\end{itemize}


\section{Background}
\label{sec:bg}

\textbf{\textsc{ethereum:}} Ethereum is a public blockchain system that uses the Turing-complete programming language Solidity and Vyper, etc to develop smart contracts\cite{li2020survey,chen2020understanding,li2020stan}. Anyone can deploy decentralized applications (dapps) on the Ethereum chain that can communicate with others, and the most popular application in the financial field is DeFi, which provides a wide range of financial services.

\textbf{\textsc{gas:}} To avoid overuse of network resources, all transactions on Ethereum are paid a cost, and the total gas cost equals the amounts of \texttt{gas} multiplied by \texttt{gasPrice} \cite{chen2017under,chen2020gaschecker}. The user who proposes transactions sets the \texttt{gasPrice}, and the transaction is conducted earlier if the \texttt{gasPrice} is high.
  
\newcommand\ChangeRT[1]{\noalign{\hrule height #1}}
  \begin{table*}[ht]
    \centering
    \caption{Smart Contract Vulnerabilities in DeFi}
    \resizebox{\linewidth}{!}{
    \renewcommand{\arraystretch}{1} 
    \begin{tabular}{cc||cc}
        \ChangeRT{0.5pt}
        \textbf{Categories} & \textbf{Causes} & \textbf{Categories} & \textbf{Causes}\\
        
        \hline
        
        Unchecked External Calls & Without checking return values &Reentry&Repeated calls before completed  \\

        Transaction State Dependency&Failure to check permissions& Nested Call& Unrestricted call depth \\
        
        DoS Under External Influence&External exceptions inside loops& Missing Return&Denote return but no value \\

        Unmatched ERC-20 Standard&Not follow the standard&Greedy Contracts & Receive but not withdraw Ethers\\

        Strict Balance Equality & Balance check failed & Block Info Dependency& Status leakage  \\

        Misleading Data Location & Incorrect \texttt{storage} type & Missing Interrupter & No backdoor to handle crises\\

        Transaction State Dependency & Error using \texttt{tx.origin} & Arithmetic Operations&Unmatched type to values\\
        
        \ChangeRT{0.5pt}
    \end{tabular}
    }
    \label{tab:smart_contract_vulner}
\end{table*}

\textbf{\textsc{miner extractable value (mev):}} It refers to the profit miners make by performing a series of operations on the blocks they mine\cite{RN7}, such as transaction inclusion, exclusion, and reordering. Miners reorder transactions to optimize the initial ordering of transactions. Earning additional ordering optimization (OO) fees\cite{RN3} is also a source of MEV.

\textbf{\textsc{development of DeFi:}} 
The introduction of blockchain technology\cite{nakamoto2008bitcoin} has changed the traditional financial ecosystem. With the advent of Ethereum, smart contracts became the basis for the development and implementation of DeFi. Since the landing of \textsc{MakerDAO} in 2014 which is the first Ethereum-based DeFi project, several DeFi protocols have emerged to implement functions of traditional CeFi, such as lending platforms, exchanges, derivatives, and margin trading systems\cite{wang2022speculative}. As liquidity mining is mentioned in 2020, DeFi is pushed into high gear with the emergence of decentralized exchanges such as \textsc{Compound}, which are entirely managed by smart contracts. \textsc{Money Legos} brings unlimited creativity to DeFi products. It means that a new financial product can be realized by combining the underlying DeFi protocols\cite{popescu2020decentralized}. In 2022, regulated decentralized finance (rDeFi) becomes the new trend in DeFi development\cite{rDeFi2022}.


\section{Analysis of Vulnerabilities}
\label{sec:ana_v}

To summarize threats in decentralized finance, we focus on data, consensus, contract, and application layers.

%
%
\subsubsection{Data Security Vulnerabilities}
\label{sec:data_security_v}
\ 
\newline
\indent\textbf{Oracle Mechanism Vulnerabilities:} The oracle is an automated service mechanism that allows the system to obtain off-chain asset data as input\cite{RN47}. However, as Fig. \ref{fig:overview} shows, the risk to oracle grows drastically when a single point of failure occurs. For example, over 3 million sETH were arbitrated (Address: \href{https://etherscan.io/tx/0x93819f6bbea390d7709fa033f5733d16418674e99c43b9ed23adb4110d657f0c}{\seqsplit{0x93819f6...}}) due to oracle errors in \textsc{Synthetix}, a protocol that converts entity into synthetic\cite{RN53}.

\indent\textbf{Inappropriate Key Management:} In the DeFi ecosystem, wallets are used to manage private keys, and authentication is based on keys in most cases. However, even the safest cryptocurrency hardware wallets have security issues\cite{RN64} caused by the architecture design. For example, the Ronin Bridge was hacked for 624 million USD, which occurred in March 2022\cite{RN23}. The hackers used a backdoor attack to get the signatures of a third-party validator and four other local verifier signatures for stealing.

\subsubsection{Consensus Mechanism Vulnerabilities}
\label{sec:con_mec_v}
\ 
\newline
\indent Certain malevolent activity leverages the rules of consensus to influence the sequences of transactions. There are a variety of attacks combined with MEV, such as flash loans\cite{qin2021attacking,RN1}, sandwich attacks\cite{RN2,RN7}, and forking attacks\cite{RN3}.

\indent\textbf{Transaction Order Vulnerability:} It will be used to describe this phenomenon in which an attacker alters the initial sequence of transactions by leveraging the miner's desire for profit. The sandwich attack is a typical example, the attacker anticipates that the victim will buy asset A, and pays a higher gas fee to miners to acquire it before the victim at a lower price. And then sells A at a higher price for arbitrage since the victim's purchase boosts the price\cite{RN2}.

\indent\textbf{Forking Vulnerability:} Forking events in DeFi are generally associated with transaction fee-based forks and time-bandit attacks\cite{RN3}. Mining revenue incentivizes miners to perform honestly, but the OO fee motivates them to reorder transactions in the block, enhancing the income.


%
%
\subsubsection{Smart Contract Vulnerabilities}
\label{sec:smart_contract_v}
\ 
\newline
\indent
There are 20 types of smart contract vulnerabilities in Ethereum defined in \cite{RN34}, of which Table \ref{tab:smart_contract_vulner} shows the security weaknesses that attackers might use to make a profit. \cite{RN41} detected over 18K real-world smart contracts and achieved an average coverage rate of 92 percent above the average of six vulnerabilities that can be categorized into the three types as detailed below:

\textbf{Suicidal and Greedy Contracts:}
Smart contracts usually include a provision enabling the owner to commit \texttt{suicide} if the contract is challenged. This suicide procedure can be carried out for any cause under the suicidal contract \cite{RN65,RN66}. Greedy contracts do not have functions related to the extraction \cite{RN65}. The contract locks all ether and cannot withdraw. Therefore, making sure there are means to get ether out before transferring it to a contract\cite{RN34}.




\textbf{Block Info Dependency:}
  In Ethereum, the discrepancy between successive blocks is valid when the timestamps are within 12 minutes\cite{RN30}. However, if the logic of the contract combines states in the block, the miner can control this information for profit\cite{RN34}. For example, when the \texttt{block.timestamp} is used as the seed in a pseudo-random function running in a contract, miners with access to this block can replicate the process of producing random numbers to attack the contract.

\textbf{Unchecked External Call:} The return value or the arguments of an external call can affect the states of the code, and many contracts do not check the return value leads to vulnerabilities. Multiple functions are nested, and the external call does not check the return value of the internal call in time can go wrong. Smart contracts in the DeFi trade by using external call functions including \texttt{call()}, \texttt{send()}, \texttt{delegatecall()}. More crucially, a failed external call in these methods results in a transaction not being rolled back, which can cause logical effects.
\subsubsection{Application Layer Vulnerabilities}
\label{sec:Dec_exchage_v}
\
\newline
\indent
The vulnerabilities at the application level are based on the manipulation of prices, and we summarize the existing flaws from the following four perspectives.

\textbf{Lending Market Imperfection:}
  When the prices in the market are out of balance, it will result in bad debts for one of the participants in the lending market. To get more loans, attackers can boost the cryptocurrency exchange rate on the oracle by modifying the real-time price-related status before the loan is made. For example, an attacker can gain a larger quantity of tokens by directly manipulating token prices in the asset pool or increasing the price of collateral before lending\cite{RN55}, putting the borrower in danger of bad debt.
  
\textbf{Cryptocurrency Instability:}
  The large fluctuations of cryptocurrencies come from many reasons, one of which is the Pump-and-Dump. The instability can easily trigger liquidation procedures. Exchanges have chosen stablecoin, which is tied to the price of real money, as the pricing standard to minimize losses, but they still exist a risk. For example, a 99.98 \%  plunge on May 11, 2022, in the price of the luna coin whose value is tied to a stablecoin called Terra, leaving the entire crypto market with over 700 million in collateral liquidated\cite{luna2022liquidation}.

  
\textbf{Design Imperfection:}
  The attackers make use of incorrectly configured functionality or specific convenience features of DeFi platform exchanges\cite{wang2021towards}. Flash loan is designed as risk-free loans to be a convenient improvement to the loan that need to borrow the flash loan, exchange it for currency and repay the loan in an atomic transaction. For example, attackers borrow the flash loan to receive collateral at a premium and make a profit in this atomic transaction \cite{RN62}, this results in bad debts for the users who borrow money from attackers.
  
\textbf{Abusive Exposure Transaction:}
  Exchanges disclose all transactions as soon as feasible to ensure complete behavioral transparency because off-chain matching services are not automated. Unfortunately, exchanges can restrict access to select users and launch denial of service attacks\cite{RN63} to dominate the market, audit transactions and even front run the orders.


\section{Analysis of attack events}
\label{sec:attack_events}
In this section, we investigate real-world attacks in DeFi and analyze the vulnerabilities exploited in the attacks.

\subsubsection{Utilization of Flash Loan}
\ 
\newline
\indent
Flash loan is a type of unsecured lending that relies on the atomicity of blockchain transactions at the point of execution\cite{qin2021attacking} and adds dynamism to DeFi. Unfortunately, attackers can exploit flaws in their existing protocols\cite{wang2021towards}. There are several other attacks caused by using the flash loan in Table \ref{tab:attack_events_flash_loan}.

From Table \ref{tab:attack_events_flash_loan}, the various attacks against the flash loan service have caused significant financial damage to the DeFi ecosystem. Attackers borrowed money from lending platforms, e.g, \textsc{dYdX}\cite{dydx}, with the flash loan services, then used the borrowed funds to manipulate the price of tokens\cite{RN55} to make an arbitrage.

Flash loans facilitate the execution and reduce the cost of attacks. The \textsc{Grim Finance} and \textsc{Popsicle Finance} incidents borrowed tokens by flash loans to enable reentry attacks and double claiming attacks, respectively.

\begin{table}[ht]
    \centering
    \caption{Attacks Related to Flash Loan }
    \setlength{\tabcolsep}{5mm}{
    \small
    \begin{tabular}{ccc}
        \ChangeRT{0.5pt}
        \makecell[c]{\textbf{Victims}} & \makecell[c]{\textbf{Date}}  & \makecell[c]{\textbf{Amount}\\(million USD)}\\
        
        \hline
        
        Harvest Finance & Oct 26, 2020 & 24  \\
        Alpha Homora & Feb 13, 2021 & 37  \\
        XToken & May 12, 2021& 24  \\
        PancakeBunny & May 19, 2021 & 200  \\
        Belt Finance & May 28, 2021 & 50  \\
        Cream Finance & Oct 27, 2021 & 130  \\
        Beanstalk Farms & Apr 18, 2022 & 182  \\
        
        \ChangeRT{0.5pt}
    \end{tabular}
    }
    \label{tab:attack_events_flash_loan}
\end{table}

\subsubsection{Private Key Leakage}
\ 
\newline
\indent
 Ethereum-based DeFi applications need to interact with the wallet, like \textsc{MetaMask}, and Ethereum provides the API \cite{ethereum-provider} that enables this interaction. Attackers get the private key of the original contract deployers or administrators to control the contract to mint or transfer tokens to others under their control. According to Table \Ref{tab:attack_events_private_key_leakage}, the exposure of the private key has lost hundreds of millions of dollars.
 \begin{table}[ht]
    \centering
    \caption{Attacks Related to Private Key Leakage}
    \setlength{\tabcolsep}{5mm}{
    \small
    \begin{tabular}{ccc}
        \ChangeRT{0.5pt}
        \makecell[c]{\textbf{Victims}} & \makecell[c]{\textbf{Date}}  & \makecell[c]{\textbf{Amount}\\(million USD)}\\
        
        \hline
        
        Meerkat Finance & Mar 04, 2021 & 31  \\
        Paid Network & Mar 05, 2021 & 160  \\
        EasyFi & Apr 19, 2021 & 80  \\
        bZx & Nov 05, 2021& 55  \\
        Vulcan Forged & Dec 13, 2021& 140  \\
        Ronin Bridge & Mar 29, 2022& 624  \\
        
        \ChangeRT{0.5pt}
    \end{tabular}
    }
    \label{tab:attack_events_private_key_leakage}
\end{table}

\subsubsection{Reentry Attack}
\ 
\newline
\indent
The most significant reentry attack in Ethereum was the DAO attack \cite{RN24} that caused a hard fork of Ethereum. Reentry attacks were applied to the DeFi protocol with its development.

The reentry attacks that occurred on the \textsc{dForce} and \textsc{Grim Finance} \cite{analysis-of-the-grim-hack} platforms, together caused a loss of 54 million USD, in Table \ref{tab:attack_events_contract_bugs}. The \textsc{dForce} incident was caused by the fact that the ERC-777 which is a standard for token contracts interfaces and behaviors allows transaction notifications to be sent to the recipient in the form of callbacks. This means that ERC-777 token indirectly results in the recipient having control of the execution\cite{RN47}.

In the \textsc{Grim Finance} security incident, the attacker publishes a malicious contract whose callback function contains a call to the \texttt{depositFor()} function in the \textsc{GrimBoostVault contract}. \texttt{depositFor()} returns proof of investment Spirit-LP to the user. Therefore, it will call the callback function in the malicious contract again to obtain multiple Spirit-LP proofs. This allows the attacker to gain more additional revenue.

\subsubsection{Arithmetic Bug}
\
\newline
\indent
Almost all DeFi applications involve arithmetic operations on currencies. These operations consist of adding or subtracting from account balances and converting exchange rates between different tokens\cite{RN47}. Attackers typically target weaknesses in arithmetic operations. This can be seen in the case of the accuracy loss in \textsc{Uranium Finance} incident, when checking the contract balance, the bug resulted in the final contract calculating 100 times larger than the actual balance\cite{uranium-finances-hacked} and losing 50 million USD.

\begin{table}[ht]
    \centering
    \caption{Attacks Related to Contract Bugs}
    \setlength{\tabcolsep}{5mm}{
    \small
    \begin{tabular}{ccc}
        \ChangeRT{0.5pt}
        \makecell[c]{\textbf{Victims}} & \makecell[c]{\textbf{Date}}  & \makecell[c]{\textbf{Amount}\\(million USD)}\\
        
        \hline
        
        dForce  & Apr 19, 2020 & 24  \\
        Uranium Finance & Apr 28, 2021 & 50  \\
        Compound & Sep 30, 2021 & 80  \\
        Grim Finance & Dec 19, 2021 & 30  \\
        
        \ChangeRT{0.5pt}
    \end{tabular}
    }
    \label{tab:attack_events_contract_bugs}
\end{table}

Another example of arithmetic vulnerabilities is the integer underflow of \textsc{Compound Finance} (Address: \href{https://etherscan.io/address/0x75442Ac771a7243433e033F3F8EaB2631e22938f#code}{\seqsplit{0x75442Ac...}}). Its reward payouts \texttt{CompSpeed} can be set to 0, which indicates that reward payouts are suspended, and the market award index \texttt{supplyIndex} is 0. For new users, their award index \texttt{supplierIndex} initialized to \texttt{CompInitialIndex} preset by \textsc{Compound} as $10^{36}$. Parameters variation causes the formula, \texttt{deltaIndex=sub\underline{ }(supplierIndex=0, supplierIndex=$10^{36}$)}, for calculating the difference in the reward index to overflow, while the calculation of the reward relies on the value of \texttt{deltaIndex}. There was no attacker in this security incident, but rather an overpayment of rewards due to an underflow vulnerability in the contract.

\subsubsection{Other Bugs}
\ 
\newline
\indent
\textbf{Attacks Related to Oracle}: Oracle serves as an information channel between the DeFi and the outside world, giving external asset values as an input source to the DeFi\cite{RN48}. \textsc{Vee Finance} requires that price variations in the mining pool of more than 3\% be re-inputted using oracle. Because its oracle solely utilizes the prices in the mining pool as an input source, the attacker can manipulate the token price in the pool, forcing the oracle to update the price.  As a result, the contract received incorrect price information, skipping the slippage protection\cite{vee-finances-hacked} and resulting in a loss of 35 million USD for \textsc{Vee Finance}.

\textbf{Phishing Attack}: DeFi website embedded scripts that we can interact with the user's wallet API, which could facilitate a phishing attack\cite{winter2021s}. The attackers used a phishing attack on BadgerDAO, causing it to lose 120 million USD. First, the attacker used the email address of the administrator to create three \textsc{Badger} accounts, one of which passed official authentication. After that, the attacker accessed the \textsc{Badger} application website through this account and injected a malicious script into the website. The script intercepted web3 transactions and prompted the user to allow the attacker to manipulate the tokens in their wallet\cite{badger}. 

\textbf{Attacks Related to Contract}: The \textsc{Wormhole} incident \cite{Wormhole} caused about 320 million USD in damage on February 3, 2022. The attacker first calls the \texttt{verify\underline{ }signature()} function to obtain signatures for the function \texttt{post\underline{ }vaa()}. However, the \texttt{load\underline{ }instruction\underline{ }at()} function called in the \texttt{verify\underline{ }signatures()} function does not verify the authenticity of the account, so the account can use the obtained valid signatures to send messages to the contract. Finally, the attacker used this vulnerability to send a message casting 120,000 wETH to the contract.

\textbf{Double-Claiming Attack}: In Table \ref{attacks related to other bugs}, the \textsc{Popsicle Finance} event\cite{begum2020blockchain} was attacked similarly to the double-spending attack which creates multiple transactions using the same cryptocurrency.

\begin{table}[ht]
    \centering
    \caption{Attacks Related to Other Bugs}
    \setlength{\tabcolsep}{5mm}{
    \small
    \begin{tabular}{ccc}
        \ChangeRT{0.5pt}
        \makecell[c]{\textbf{Victims}} & \makecell[c]{\textbf{Date}}  & \makecell[c]{\textbf{Amount}\\(million USD)}\\
        
        \hline
        
        Spartan Protocol & May 02, 2021 & 30  \\
        Popsicle Finance & Aug 03, 2021 & 25  \\
        Poly Network & Aug 10, 2021 & 26  \\
        Vee Finance & Sep 21, 2021 & 37  \\
        BadgerDAO & Dec 02, 2021 & 120  \\
        Qubit Finance & Jan 28, 2022 & 80  \\
        Wormhole & Feb 03, 2022 & 326  \\
        
        \ChangeRT{0.5pt}
    \end{tabular}
    }
    \label{attacks related to other bugs}
\end{table}

First, the attacker deposits funds via \textsc{Popsicle Finance}, and the platform returns a PLP Token certificate of deposit. Then, the attacker transfers the certificate to other contracts under his control. \textsc{Popsicle Finance} calculates the user's reward incrementally via the \texttt{\underline{ }fee0Earned()} function. The rewards are accumulated even if there is no asset in the user's account. Finally, the attacker controls the contract by calling the \texttt{withdraw()} function to remove the deposited funds and rewards.

\section{Analysis of Security optimization}

\subsubsection{Data Security Optimization}
\ 
\newline
\indent
\textbf{Oracle optimization Schemes:} Due to the necessity for off-chain asset information such as pricing, as discussed in \ref{sec:data_security_v}, there is an expanding demand for superior oracles\cite{RN68}. Our research looked at the real-world oracle optimization choices of DeFi systems. The \textsc{Compound} that aggregates pricing from off-chain to on-chain via the \textsc{ChainLink} \cite{RN48}, delivers multi-party data directly to the contract through reputation from providers, forming a reference pricing network where nodes in the chain may get price data to stay up to current. However, quantitative reputation cannot match the oversized price makes it can only apply on a small scale. Another form is \textsc{MakerDAO} \cite{Maker_Protocol}, which collects off-chain data through the central medianizer which is an aggregator. It utilizes the median of prices for pricing and delays price updates by one hour before uploading on the chain so that governors and users can react to faults to secure the process.

\textbf{Wallet Key Security Optimization:} Users initiate a transaction and sign it using the key pair, the assets in the account are lost when the key leaks to an adversary. Some studies\cite{RN74,RN77,RN80} proposed specific solutions for wallet management and wallet architecture.
  According to\cite{RN64}, existing hardware wallets migrated from the PC wallet architecture, resulting in a bad design that does not fundamentally fix the problem when just utilizing authentication and communication encryption. For interactive authentication, adds several signatures and keys to the original wallet structure, which prevents attackers from manipulating the keys for transactions using a malfunctioning wallet.
  Combined with software and hardware, two android applications created in\cite{RN80} for a cold wallet with key storage in the form of QR codes and a hot wallet for sending transactions, respectively, provide privacy protection.

\subsubsection{Smart Contract Security Optimization}
\
\newline
\indent
The smart contract, which is a part of the DeFi project connecting the data and the application layer, might alter the state of a transaction, and cause errors, so it's critical to improve the security of contracts.

\begin{figure}[htp]
    \centering
    \includegraphics[width=8cm]{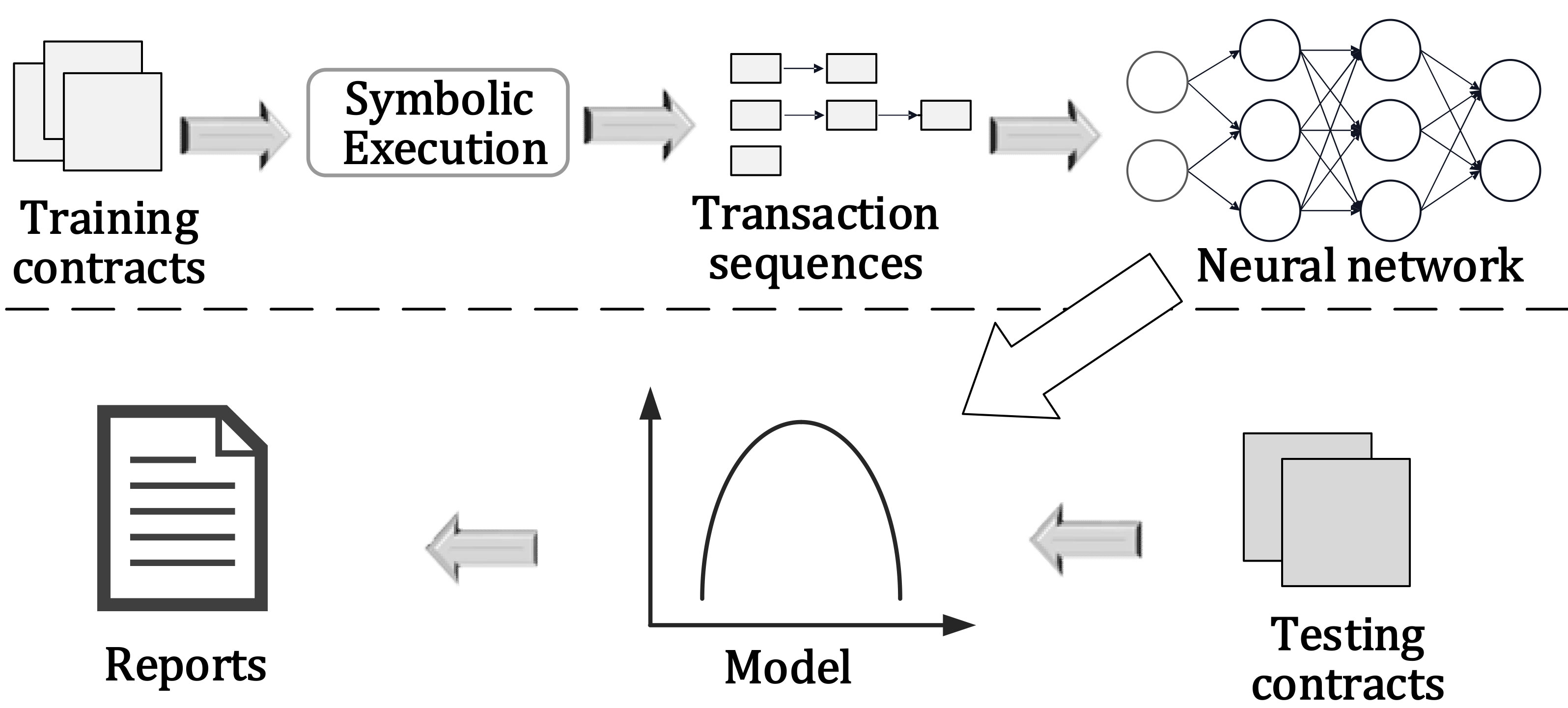}
    \caption{Schematic Diagram of \textsc{ILF} Process Framework}
    \label{fig:ILF}
\end{figure}

\textbf{Smart Contract Vulnerability Detection:} Much research\cite{RN26,RN28,RN39,RN41,RN81} has been undertaken to discover contract vulnerabilities using various technological tools, such as formal verification, and machine learning. Combined with dynamic testing extends the ability of symbolic execution techniques to detect unknown vulnerabilities, thus improving the robustness of programs. Fig. \ref{fig:ILF} shows an overview of \textsc{ILF}~\cite{RN39} that combines fuzzing, machine learning, and symbolic execution.

The system used the symbolic execution for a portion of the contracts to generate transaction sequences as the training dataset for a new model consisting of GRU which is a type of neural network and a fully connected network so that the model can learn the fuzzing in the state after the symbolic execution to test contracts with high coverage.

\begin{figure}[htp]
    \centering
    \includegraphics[width=6cm]{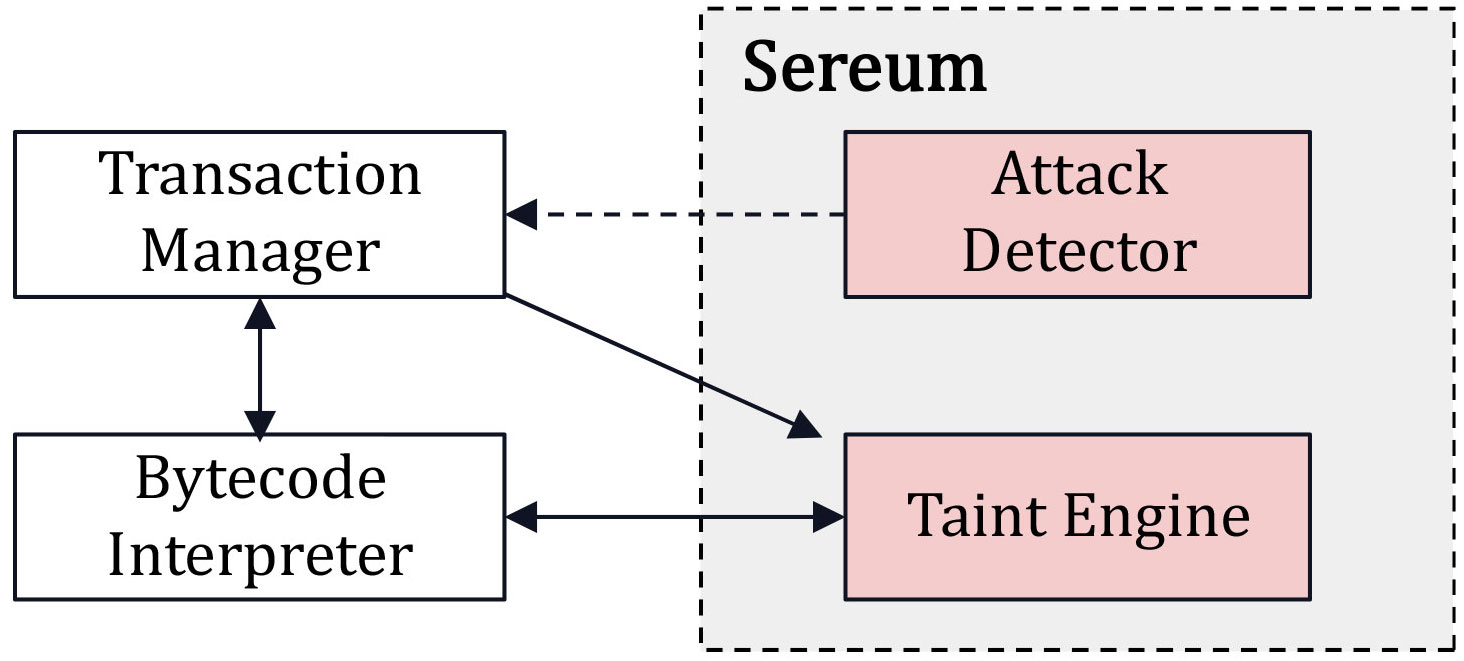}
    \caption{Diagram of \textsc{Sereum} System Architecture}
    \label{fig:Sereum}
\end{figure}

\textbf{Smart Contract Operation Regulation:}
However, \cite{RN87} showed that contract vulnerability detection can increase contract defensibility, but assaults have not decreased, indicating that contract regulation has to be further improved. It has been studied in\cite{RN37,RN87,RN90,RN91,RN92,RN93}, and we briefly introduce \textsc{Sereum}~\cite{RN93} in Fig. \ref{fig:Sereum}, a security tool focused on runtime monitoring and verification of reentry vulnerability.

Transaction Manager converts all control flows into conditional jump instructions in the bytecode interpreter, and then the taint engine identifies data flows in conditional jump instructions, tagging \texttt{storage} variables as key variables and writing into the lock. The attack detector detects the variables, if the modification occurs, the whole transaction is rolled back to the point where the variable was marked, which is the starting point of the entire transaction.

\subsubsection{Consensus Layer Optimization}
\
\newline
\indent
The consensus layer and the incentive layer are interdependent, and the design of the consensus mechanism directly affects the behavior of miners, although many consensus mechanisms have been proposed, there is little regulation of the consensus and incentive levels.

\begin{figure}[htp]
    \centering
    \includegraphics[width=8cm]{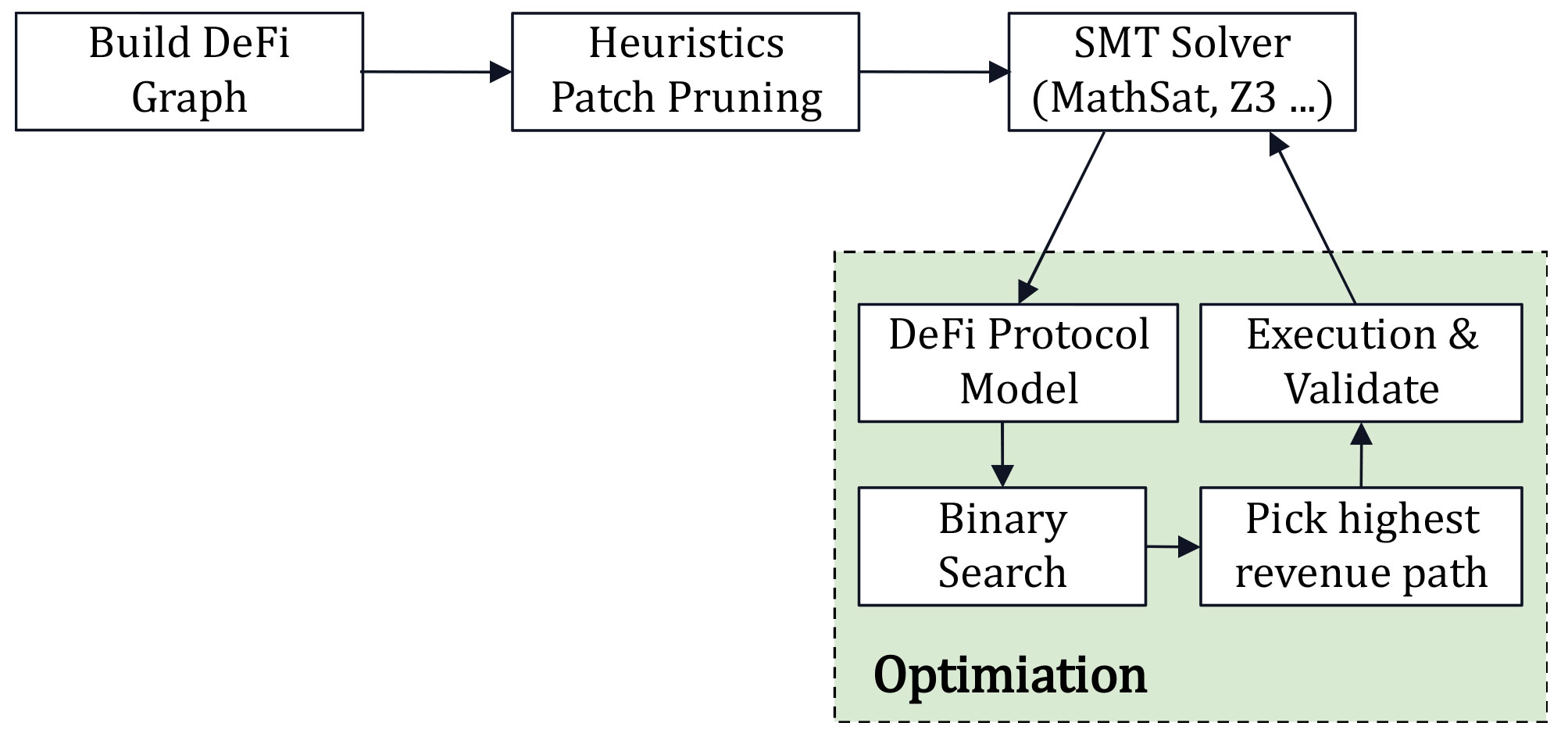}
    \caption{Diagram of \textsc{DefiPoser} Core Process}
    \label{fig:DefiPoser}
\end{figure}

As described in \ref{sec:con_mec_v}, fork attacks might affect blockchain security in terms of consensus mechanism, \cite{RN1} developed \textsc{DefiPoser} to monitor fork behaviors. Fig. \ref{fig:DefiPoser} shows the process of \textsc{DefiPoser}, it heuristically prunes the patches after building the DeFi graph and then does a greedy search of the negative cycle in the directed transaction flow graph, which means finding all possible profitable cycles in the trade flow graph, to detect arbitrage transactions in cyclic or more complicated scenarios. A binary search of all the paths finds the most profitable one. If it is within the quantization threshold quantified by the Markov decision process, there is an opportunity to motivate a fork attack by miners using MEV.

\subsubsection{P2P Network Optimization}
\
\newline
\indent
The transactions initiated by each node in Ethereum are transmitted through P2P networks to achieve self-governing without relying on a third party; however, the lack of authentication and other features leads to a series of attacks, such as the eclipse attack\cite{RN103,RN102,RN105} and sybil attack\cite{RN104}. An information eclipse attack occurs when an aggressor removes nodes from a network to restrict access to information from nodes.

However, \cite{RN102} suggests a series of protection methods against eclipse attacks on the Ethereum, two of which are also adopted by \texttt{geth}. When a node restarts, the client's \texttt{seeding} is triggered every hour, or \texttt{lookup()} is called on an empty \texttt{table} which stores the information in \texttt{memory}, but the \texttt{seeding} is available only if the \texttt{table} is empty. However, node IDs should always be inserted into the \texttt{table} to prevent attacks. Specifically, \texttt{geth} runs a \texttt{lookup()} on three random targets during \texttt{seeding} to add more legitimate nodes from the \texttt{db} which stores the information on disk to the \texttt{table} to prevent attackers from inserting their node IDs into an empty \texttt{table} during \texttt{seeding}.

\subsubsection{Application Layer Optimization}
\
\newline
\indent
Although there is a correlation between the various layers, methods for lower levels can not fully recognize the attacks against the application layer. There still exists some research\cite{RN55,RN90} that makes contributions.

\begin{figure}[htp]
    \centering
    \includegraphics[width=8cm]{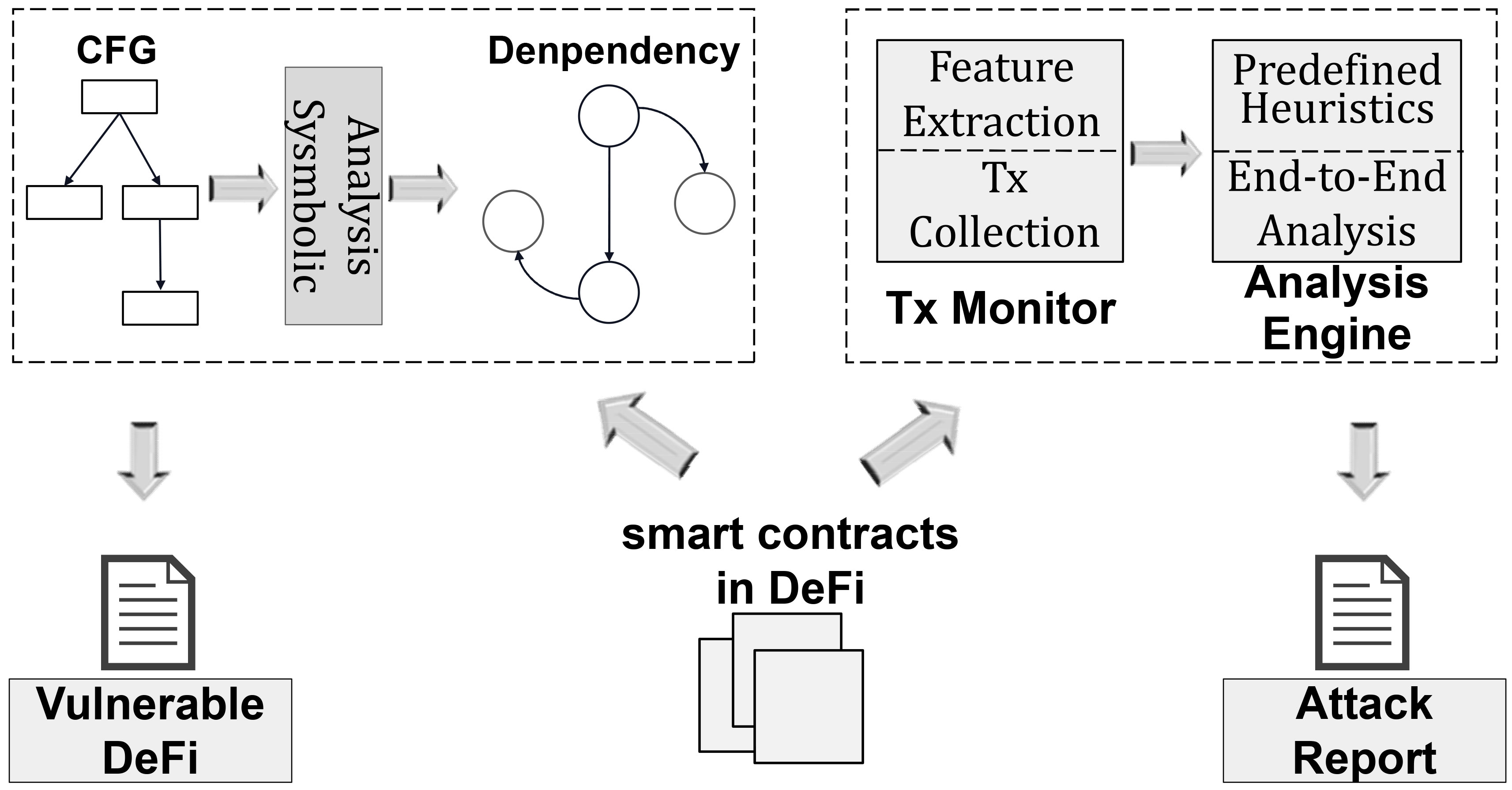}
    \caption{Diagram of \textsc{Blockeye} Core Process}
    \label{fig:Blockeye}
\end{figure}

\cite{RN90} designed \textsc{BLOCKEYE} divides the detection work into two phases. In Fig. \ref{fig:Blockeye}, the first phase uses symbolic execution analysis in oracle to check whether state data streams are externally manipulated to detect vulnerable DeFi, and during the second phase, transaction monitors under the chain collect transactions to extract the features and further analysis to monitor the attack.

\subsubsection{Insurance Optimization}
\ 
\newline
\indent
As the DeFi market expands, the insurance on it is critical to ensuring its stability\cite{RN108}. Our research divides risks in DeFi into market risks, technical risks, and credit risks. However, the damages experienced by regular users as a result of technical or credit risks are enormous, and an insurance system is required to safeguard the properties of users. It can classify as centralized and decentralized.

For example, \textsc{Opyn}\cite{RN109}, which focuses on insurance for option trading products, enables users to choose options to hedge risks based on ERC20 tokens, and the protocol is automatically performed by smart contracts for multiparty governance. \textsc{Smart Contract Cover}, which provides smart contract insurance, is evaluated by the company \textsc{Nexus Mutual} \cite{RN108} internal assessor to determine the cost of the insurance.



\section{Conclusion and Future Direction}


The focus of this paper is on the security of DeFi, and we summarize a series of security risks of DeFi by analyzing their projects deployed in Ethereum. For each vulnerability, we explore its causes with real-world cases. Finally, we investigate the optimization options for decentralized finance and suggest possible future directions.

Comprehensive knowledge of security and risk problems is critical to improving blockchain and establishing powerful defense capabilities in practice. There is a strong possibility to combine static detection with dynamic supervision technologies to protect DeFi at the consensus mechanism, smart contract, and application levels for the future development of DeFi application security.


%
\IEEEpeerreviewmaketitle






\normalem
\bibliographystyle{IEEEtran}
\bibliography{ref}
\end{document}